\documentclass[11pt]{article} 
\usepackage{amsfonts,amscd,amsmath,graphicx}

\def\NP#1#2{ Nucl.Phys. B#1 (#2)} 
\def\PL#1#2{ Phys.Lett. B#1 (#2)}
\def\CMP#1#2{ Commun.Math.Phys. #1 (#2)}

\def\PR#1#2{Phys.Rev. D#1 (#2)} 
\def\IJMP#1#2{ Int.J.Mod.Phys. A#1 (#2)}
\def\ATMP#1#2{ Adv.Theor.Math.Phys. #1 (#2)}
\def\HP#1#2{ JHEP #1 (#2)} 

\newcommand{\so}{\hat{sl}(2)}
\newcommand{\ssl}{\hat{sl}(2\vert 1)} 
\newcommand{\su}{\hat{su}(2)}
\newcommand{\zo}{z_{12}} 
\newcommand{\zt}{z_{12}^2}
\newcommand{\pd}{\partial} 
\newcommand{\oh}{\frac{1}{2}}
\newcommand{\oq}{\frac{1}{4}} 
              
\newcommand{\pp}{\dot +} 
\newcommand{\mm}{\dot -}
\newcommand{\te}{\tilde\eta} 
\newcommand{\txi}{\tilde\xi}
\newcommand{\bz}{\bar z} 
\newcommand{\sif}{\sigma^1}
\newcommand{\sis}{\sigma^2} 
\newcommand{\ep}{\text e}

\newcommand{\ads}{\text{AdS}_3}
\newcommand{\vaf}{\varphi^1}
\newcommand{\vas}{\varphi^2} 

\title{Unitary Representations of Some Infinite Dimensional Lie
  Algebras Motivated by String Theory on $\ads$}
\author{Oleg Andreev\thanks{E-mail: andre@landau.ac.ru}\\ \\
  Landau Institute for Theoretical Physics,\\
  Kosygina 2, 117334 Moscow, Russia} \date{}
\begin{document}
\maketitle

\vspace{-8cm}
\begin{flushright}hep-th/9905002 \\
  LANDAU-99/HEP-A2
\end{flushright}

\vspace{6cm}
\begin{abstract}
  We consider some unitary representations of infinite dimensional Lie
  algebras motivated by string theory on $\ads$.  These include
  examples of two kinds: the A,D,E type affine Lie algebras and the
  $N=4$ superconformal algebra. The first presents a new construction
  for free field representations of affine Lie algebras. The second
  is of a particular physical interest because it provides some hints
  that a hybrid of the NSR
  and GS formulations for string theory on $\ads$ exists. 
\end{abstract}

\section{Introduction} 
\renewcommand{\theequation}{1.\arabic{equation}}
\setcounter{equation}{0}

This is our second paper in a series of papers in which several
physical ideas, (i) the 't Hooft Holographic principle \cite{H,LS}, in
particular the AdS/CFT correspondence \cite{Mal}; (ii) the Green's
world-sheets for worlds-sheets \cite{Green}; (iii) topological quantum
numbers as central charges of superalgebras \cite{SUSY}, are used as
the keystones upon which better understanding of superstring theory on
some curved space-times, specifically $\ads$, should be achieved. The
possible relation of string theory on $\ads$ to two-dimensional
conformal field theory has been a subject of much interest (see, e.g.,
[7-18] and references therein\footnote{It should be noted that the issue 
of the string propagation on $\ads$ is an old story (for earlier 
discussions of this issue see \cite{old} and references therein).}) 
and has led to a proposal for new 
representations of generators of some Kac-Moody and superconformal
(superVirasoro) algebras \cite{GKS,I,A}. Thanks to the underlying
world-sheet structure such construction may be called the stringy
representation of affine algebras (superalgebras). So far, however,
the missing point of these stringy realizations of the algebras has
been their representations. Since we are interested in physics it is
natural to restrict ourselves to a particular class of
representations, namely the unitary highest weight representations.
According to \cite{BPZ}, the Hilbert space of any two-dimensional
conformal field theory decomposes into holomorphic and
anti-holomorphic sectors, so it makes sense to start with considering
one of them, say holomorphic. Moreover, the holomorphic sector itself
is a direct product of representations.  Thus our purpose is to build
the representations in certain two-dimensional constructions that are
of a particular physical interest. However, before we give examples it
might be useful to present some mathematical background.

In general, given a finite-dimensional Lie algebra $\mathfrak g$, the
untwisted affine Kac-Moody algebra $\hat\mathfrak g$ is simply defined
by $\hat\mathfrak g =\mathfrak
g\otimes\mathbf{C}[\gamma,\,\gamma^{-1}]\oplus
\mathbf{C}k\oplus\mathbf{C}d$, where $k$ is a central element and $d$
is a derivation \cite{Kac}. There are several candidates for what the
parameter $\gamma$ can be. Since we are interested in 2d conformal
field theory, our candidate is a coordinate on a Riemann surface.
Indeed, this is exactly what we need to build conformal field theory
from representations of the Virasoro (Kac-Moody) algebra, namely to
define a construction attaching a representation of $\hat\mathfrak g$
to a point on the surface. 

This is an old story, however. Now, let us assume that there is a new
Riemann surface whose coordinate is $z$ such that $\gamma$ is a
function of $z$. Such an assumption arises naturally if one thinks of
the second Riemann surface as the world-sheet for the first one, i.e.
these are the Green's world-sheets for world-sheets. More rigorously,
let $P_0$ be a point on the world-sheet which is a Riemann surface and
$(U_0,z)$ be a coordinate patch such that $P_0\in U_0$ and $z=0$ for
$P_0$. Let $P_\gamma$ be a point on another two-dimensional manifold
(space-time) and $(U_\gamma,\gamma)$ be a proper coordinate patch. Let
there be a map $\gamma :\,\, U_0\rightarrow U_\gamma$, i.e. a local
embedding of the world-sheet into the space-time. It is useful to
classify such maps by their degrees, i.e.  integers $k$ given by
\begin{equation}\label{top}
  k=\oint_{C_0} dz\,\pd\gamma\gamma^{-1}\,(z)
  \quad,
\end{equation}
where the contour $C_0$ surrounds $0$. Thanks to the local
construction $C_0$ is contractible on the world-sheet. The
normalization is fixed by setting $k=1$ for $\gamma (z)=z$ i.e.,
$\oint_{C_0}\frac{dz}{z}=1$. In the quantum case $k$ will be
interpreted as the central (topological) charge\footnote{Note that in
  the stringy context $k$ is interpreted as the number of infinitely
  stretched fundamental strings at the boundary of $\ads$ while in the
  framework of the $(\beta,\gamma)$ system $-k$ is called the
  ``Bose-sea level''.}.

The outline of the paper is as follows. As a warmup, we start in
section 2 by describing how a simple set of free fields on the
world-sheet, namely a bosonic first order system together with a
scalar field provides the $\su$ algebra in the space-time. We show
that it is natural to consider two types of representations of this
algebra. The first type is generated by string vertices i.e., vertex
operators whose conformal dimensions in $z$ are $1$, integrated over
$dz$. We call these as the stringy representations. The second type is
similar to the representations known from 2d conformal field theory.
They are generated by vertex operators whose conformal dimensions in
$z$ are arbitrary. So, we call them as the CFT representations. Both
constructions lead to the unitary representations of $\su$.

We then go on in section 3 to generalize the results of section 2 for
any affine Lie algebra $\hat\mathfrak{g}$ of A, D, E type. This is
done by simply adding more scalar fields such that their number equals
rank$\mathfrak{g}$.  On the one hand, this construction is reminiscent
of the vertex operator construction due to the same number of the
scalar fields. On the other hand, it is similar to the Wakimoto free
field representation due to the presence of the bosonic first order
system. However, this is, in fact, a new construction for free field
representations of affine Lie algebras. As far as we know, it has not
been studied in the mathematical literature yet.

In section 4, we generalize the construction of section 2 by adding a
new scalar field with a proper background charge. In this formulation,
we show that on the one hand, the space-time theory possesses the
$N=4$ superconformal algebra but, on the other hand, the underlying
world-sheet theory possesses the $\so\times\su$ algebra of level 1.
This is an interesting phenomenon that provides evidence for a hybrid
of the NSR and GS formulations for string propagation on $\ads$. In
fact, starting with the NSR formalism assuming all machinery of 2d
CFT, we do not need the $N=1$ structure, i.e. superconformal symmetry,
on the world-sheet to get the manifest space-time supersymmetry!

Finally, section 5 will present our conclusions and directions for
future work.


\section{ The SU(2) case} 
\renewcommand{\theequation}{2.\arabic{equation}}
\setcounter{equation}{0}

Let us consider first the $\su$ algebra in the space-time.  For a
quantum field $\gamma(z)$ which classically defines the embedding of
the world-sheet into the space-time, we introduce its conjugate field
$\beta(z)$ so that these define a first order $(\beta,\gamma)$ system
of weight $(1,0)$. Moreover, let us assume that there is one more free
field on the world-sheet. For simplicity, let it be a scalar field
$\varphi(z)$ without a background charge, which provides the $c=1$
theory. It is well-known that its Fock spaces are the representations
of the $\su$ affine Lie algebra at the level $k=1$. To be precise, in
the Fock spaces the corresponding holomorphic currents (generation
functions of $\su$ generators) are represented by
\begin{equation}\label{su-2}
  H(z)=i\pd\varphi (z)
  \quad,\quad
  E^{\pm}(z)=\ep^{\pm i\sqrt 2\varphi(z)}
  \quad.
\end{equation}
This is the simplest case of the vertex operator construction, by
which a simply-laced affine Lie algebra at the level $1$ can be
realized in terms of free scalar fields without background charges
\cite{Hal,vertex}.

A possible way to formulate the space-time $\su$ algebra is to define
its generators as \cite{GKS}\footnote{We use bold letters for
  space-time generators here and below. }
\begin{equation}\label{su2}
  \mathbf{H}_n=\oint_{C_0} dz\,\gamma^n H(z)
  \quad,\quad 
  \mathbf{E}_n^{\pm}=\oint_{C_0} dz\,\gamma^n E^{\pm}(z)
  \quad,\quad 
  n\in\mathbf{Z}\quad.
\end{equation}
For $\gamma(z)=const$, this definition reduces to
$\mathbf{H}_n=H_0\gamma^n\,,\,\,\mathbf{E}^{\pm}_n=E_0^{\pm}\gamma^n$
that agrees, of course, with the definition of the generators for the
centerless $\su$ algebra \cite{Kac}.

Conformal techniques may be used to calculate the commutation
relations. By using the OP expansions
$\varphi(z_1)\varphi(z_2)=-\ln\zo+O(1)$ and
$\beta(z_1)\gamma(z_2)=\frac{1}{\zo }+O(1)$ for $\vert z_1\vert >\vert
z_2\vert$, one finds
\begin{equation}\label{su2-st}
  [\mathbf{H}_n,\mathbf{H}_m]=\mathbf{k}n\delta_{n+m,0}
  \quad,\quad
  [\mathbf{H}_n,\mathbf{E}^{\pm}_m]=\pm\sqrt 2\mathbf{E}^{\pm}_{m+n}
  \quad,\quad
  [\mathbf{E}^+_n,\mathbf{E}^-_m]=\mathbf{k}n\delta_{n+m,0}+
  \sqrt2\mathbf{H}_{n+m}
  \quad
\end{equation}
together with
\begin{equation}
  [\mathbf{H}_n,\mathbf{k}]=[\mathbf{E}^{\pm}_n,\mathbf{k}]=0
  \quad. 
\end{equation}
Here the central element $\mathbf{k}$ is defined by \eqref{top}, but
the field $\gamma$ is now quantized.

However, this cannot be the whole story, for the reason indicated in
the introduction. It is well-known that the $\su$ algebra possesses
the unitary highest weight representations that are defined as
follows.  Let $\vert\mu\rangle$ be the highest weight vector of $\su$
with the weight $\mu$ namely,
\begin{equation}\label{hwv}
  \mathbf{H}_0\vert\mu\rangle=\mu\vert\mu\rangle
  \quad,\quad
  \mathbf{k}\vert\mu\rangle=k\vert\mu\rangle
  \quad,\quad
  \mathbf{E}^+_0\vert\mu\rangle=0
  \quad,\quad
  \mathbf{H}_n\vert\mu\rangle=\mathbf{E}^{\pm}_n\vert\mu\rangle=0
  \quad\text{for}\quad n>0
  \quad.
\end{equation}
The representation $V_{\mu}$ is built as the Verma module over $\su$
generated by the vector $\vert\mu\rangle$. Note that the unitary
representations only exist for
\begin{equation}\label{su2-u}
  k\in\mathbf{N}\quad,\quad
  0\leq\sqrt2\mu\leq k\quad\text{with}\quad\sqrt 2\mu\in\mathbf{N}
  \quad.
\end{equation}
Since we are interested in stringy applications it is natural to look
for the unitary representations. One more reason which forces us to do
this is the integer level $k$. For such values, other degenerate
representations (admissible reps.) do not appear.

For reasons that are standard and will become clear latter, it is
convenient to first bosonize the $(\beta,\gamma)$ system. The general
recipe of \cite{FMS} tells us to set
\begin{equation}\label{bos}
  \gamma(z)=\ep^{i\sif (z) -\sis (z)}
  \quad,\quad
  \beta(z)=i\pd\sif\ep^{-i\sif (z)+\sis (z)}
  \quad,
\end{equation}
where $\sif ,\,\sis$ are the scalar fields with proper background
charges and two-point functions normalized as
$\langle\sigma^i(z_1)\sigma^j(z_2)\rangle= -\delta^{ij}\ln\zo$.

In terms of these scalar fields, generators \eqref{su2} take the form
\begin{equation}\label{su2b}
  \mathbf{H}_n=\oint_{C_0} dz\,\ep^{n(i\sif -\sis )}H(z)
  \quad,\quad 
  \mathbf{E}_n^{\pm}=\oint_{C_0} dz\,\ep^{n(i\sif -\sis )}E^{\pm}(z)
  \quad.
\end{equation}
Moreover, the formula for the central element $\mathbf{k}$ becomes
\begin{equation}\label{kb}
  \mathbf{k}=\oint_{C_0} dz\,i\pd\sif -\pd\sis (z)
  \quad.
\end{equation}

\subsection{ Stringy representations}

In general, the integrands of the string vertex operators
corresponding to the physical states are primary conformal fields.
Moreover, the reparametrization (conformal) invariance of the theory
dictates that such primaries have dimension one in both $z$ and $\bz$
so that their integral over $dzd\bz$ is invariant (see e.g.,
\cite{FMS}). This definition or more exactly its holomorphic version
is equivalent to
\begin{equation}\label{phys}
  [L_n,{\cal O}]=0
  \quad,
\end{equation}
where ${\cal O}$ is a string vertex operator and $L_n$ are the
Virasoro generators on the world-sheet i.e.,
\begin{equation}\label{vir}
  L_n=\oint_{C_0} dz\,z^{n+1}\bigl(\beta\pd\gamma-
  \oh\pd\varphi\pd\varphi\bigr)(z)
  \quad.
\end{equation}
As a small technical matter, one can check that the generators
\eqref{su2} satisfy this requirement. So they are the string vertex
operators.

Looking for a highest weight vector of the algebra $\su$, we see that
there is a natural possibility:
\begin{equation}\label{r-hwv}
  {\cal O}^\mu=\oint_{C_0} dz\,\ep^{ix^+\sif +x^-\sis }\,
  \ep^{ix\varphi (z)}
  \quad.
\end{equation}
In fact, the action of $\mathbf{H}_n,\,\mathbf{E}^{\pm}_n,\,L_n$ on
string vertex operators of this type is defined by
\begin{equation}\label{act}
  T\bigl({\cal O}\bigr)=\oint_{C_0}dz_2\oint_{C_{z_2}}dz_1\,
  J(z_1)V(z_2)
  \quad\text{with}\quad
  {\cal O}=\oint_{C_0}dz\,V(z)
  \,\,,\,\,
  T=\oint_{C_0}dz\,J(z)
  \,\,,\,\,
  T=\mathbf{H}_n,\mathbf{E}^{\pm}_n,L_n
  \,\,.
\end{equation}
The parameters $x^{\pm},\,x^0$ are easily determined. According to
\eqref{hwv}, the action of $\mathbf{H}_0$ on ${\cal O}^\mu$ is given
by $\mu{\cal O}^\mu$, so $x=\mu$. Similarly, $\mathbf{k}$ acts as
$k{\cal O}^\mu$, which results in $x^++x^-=k$. Moreover, we know that
the integrand of the string vertex is the conformal primary of
dimension $1$ or, equivalently, it must obey \eqref{phys}. This is
true for $x^{\pm}$ such that ${\mu}^2+k(x^+-x^-+1)=2$. Thus the
allowed values of $x^{\pm}$ are $\oh(k\pm\frac{1}{k}(2-k-\mu^2))$,
respectively.

The above definition can also be written as
\begin{equation}\label{act1}
  T\bigl({\cal O}\bigr)=[T,{\cal O}]
\end{equation}
by rescaling ${\cal O}^\mu\rightarrow {\cal O}^\mu c_\mu$, where
$c_\mu$ is a proper cocycle factor. It is explicitly given by
$c_\mu=(-)^{\mu p+x^+\rho^1-x^-\rho^2}$ with
$p=i\oint_{C_0}dz\,\pd\varphi(z)\,,\,\,\rho^1=i\oint_{C_0}dz\,\pd\sif
(z)\,, \,\,\rho^2=-\oint_{C_0}dz\,\pd\sis (z)$.

Before going on, it is interesting to look at the structure of the
integrand for the above vertex operator. The operator
$\ep^{i\mu\varphi (z)}$ turns out to be the highest weight vector for
the $su(2)$ multiplet of the $c=1$ theory, while the remaining
operator ${\cal V}_k(z)=\ep^{ix^+\sif +x^-\sis (z)}$ can be
interpreted as an operator that interpolates between the various Bose
sea levels of the $(\beta,\gamma)$ system, shifting the level by
$-k\,\,$\footnote{For $x^+\not=0$, actually, this does not create an
  exact vacuum of the $(\beta,\gamma)$ system with the Bose sea-level
  $-k$ from the vacuum whose Bose sea-level is zero. We will return to
  a discussion of this point in the next subsection.}.  On the other
hand, it makes sense to speak of the level $1$ $SU(2)$ conformal field
theory dressed by the $(\beta,\gamma)$ system due to some analogy with
2d gravity or, equivalently, $1+1$ dimensional string theory.
Effectively, this $SU(2)$ theory has two primary fields corresponding to
the highest weight vectors namely, $1$ and $\ep^{i/\sqrt 2\varphi(z)}$ 
because $\sqrt 2\mu\leq 1$. Therefore, it is at first sight natural to expect 
that the primaries of the space-time theory are these $SU(2)$ fields 
dressed by proper $(\beta,\gamma)$ fields. We will show a
  little bit later that the allowed values of $\mu$ are $0,\,1/\sqrt
  2, \dots,\, k/\sqrt 2$. This means that we need fields that are outside the 
basic grid of the level $1$ $SU(2)$ conformal field theory. It is well-known 
that the same phenomenon happens in 2d gravity coupled to the minimal 
conformal matter.

Now, we come to more a detailed analysis of \eqref{r-hwv} by using
\eqref{act}.  For $\mathbf{H}_n$, we get
\begin{equation}\label{act0}
  \mathbf{H}_n\bigl({\cal O}^\mu\bigr)=\oint_{C_0}dz_2
  \oint_{C_{z_2}}dz_1\,(\zo )^{nk-1}\Bigl(\mu +O(\zo )\Bigr)
  \ep^{n(i\sif -\sis )(z_1)}{\cal V}_k\ep^{i\mu\varphi (z_2)}
  \quad.
\end{equation}
{}From \eqref{act0}, we see that $k$ must be integer. Otherwise the
contour integrals would be ill-defined. Moreover, $k$ must be positive
for ${\cal O}^\mu$ to be annihilated by $\mathbf{H}_n$ with $n>0$.
In particular, $\mathbf{H}_0$ acts as claimed before.

Similarly, $\mathbf{E}^{\pm}_n$ act as
\begin{equation}\label{act+-}
  \mathbf{E}^{\pm}_n\bigl({\cal O}^\mu\bigr)=\oint_{C_0}dz_2
  \oint_{C_{z_2}}dz_1\,(\zo )^{nk\pm\sqrt2\mu}
  \ep^{\pm i\sqrt 2\varphi}\ep^{n(i\sif -\sis )(z_1)}
  {\cal V}_k\ep^{i\mu\varphi (z_2)}
  \quad.
\end{equation}
First of all let us note that the contour integrals are well-defined
only for $\sqrt 2\mu\in\mathbf{Z}$. By considering the
$\mathbf{E}^+_n$ with $n\geq 0$, we see that all of the integrands are
non-singular for $\mu\geq 0$. Thus these generators annihilate ${\cal O}^\mu$. 
On the other hand, negative powers of $\zo$ may arise for $\mathbf{E}^-_n$
with $n>0$ unless $\sqrt 2\mu\leq k$.  Therefore, the highest weight
vectors exist only for $\mu=0,\,1/\sqrt 2,\dots ,k/\sqrt 2$, which is
of course the desired result because exactly such values of $\mu$ and
$k$ define the unitary representations (see \eqref{su2-u}).

Thus we have built the highest weight vectors for the $\su$ algebra
defined by \eqref{su2}. It remains to show that the Verma modules
generated by such vectors contain only the string vertex operators.
This can be done inductively, using the fact that $\mathbf{H}_n$,
$\mathbf{E}^{\pm}_n$ and ${\cal O}^\mu$ are stringy. If we take, for
example, $\mathbf{H}_n\bigl({\cal O}^\mu\bigr)$ then
\begin{equation}\label{ind}
  L_m\mathbf{H}_n\bigl({\cal O}^\mu\bigr)=
  \mathbf{H}_nL_m\bigl({\cal O}^\mu\bigr)=0
\end{equation}
or taking the cocycle factor into account,
\begin{equation}\label{ind1}
  [L_m,\mathbf{H}_n\bigl({\cal O}^\mu\bigr)]
  =[L_m,[\mathbf{H}_n,{\cal O}^\mu]]=
  [\mathbf{H}_n,[L_m,{\cal O}^\mu]]-
  [{\cal O}^\mu,[L_m,\mathbf{H}_n]]=0
  \quad.
\end{equation}
The same can be easily repeated for a general state from the Verma
module. So we have built the unitary highest weight representations of
$\su$. At this point, a couple of comments are in order.

(i) In addition to the highest weight vectors built above, there are
other vectors that may be important in the future.  First let us
determine the string vertices corresponding to the lowest weight
vectors for the $su(2)$ multiplets of the $c=1$ theory. They are
related to the highest weight vectors as ${\cal O}^{-\mu}=
\bigl(\mathbf{E}^-_0\bigr)^{\sqrt 2\mu}\bigl({\cal O}^\mu\bigr)$.
After a simple calculation, we find
\begin{equation}\label{lwv}
  {\cal O}^{-\mu}\sim
  \oint_{C_0} dz\,{\cal V}_k\ep^{-i\mu\varphi (z)}
  \quad.
\end{equation}
Now we want to continue calculations to determine the stringy vertices
corresponding to some of the singular vectors of $V_\mu$. The singular
vectors of interest are $\bigl(\mathbf{E}^-_0\bigr)^{\sqrt
  2\mu+1}\vert\mu\rangle$ and $\bigl(\mathbf{E}^+_{-1}\bigr)^{k-\sqrt
  2\mu+1}\vert\mu\rangle$.  We see from \eqref{act+-} and \eqref{lwv}
that there is no short distance singularity in the integrand
corresponding to $\bigl(\mathbf{E}^-_0\bigr)^{\sqrt 2\mu+1}\bigl({\cal
  O}^\mu\bigr)$.  Thus, it must vanish for $\mu\geq 0$. Similarly, a
relatively long but straightforward calculation shows that
$\bigl(\mathbf{E}^+_{-1}\bigr)^{k-\sqrt 2\mu+1}\bigl({\cal
  O}^\mu\bigr) =0$ for $\sqrt 2\mu\leq k$. 

(ii) Having the representation $V_\mu$ of the space-time $\su$
algebra, it is worth saying about the space-time meaning of such
states. In fact, the dictionary for comparing the conformal family
(fields) to the space of states (representation) is known \cite{BPZ}.
Let us give an example of the use of this dictionary. For the highest
weight vector ${\cal O}^\mu$, we have ${\cal O}^\mu
=\lim_{\gamma\rightarrow 0}\Phi^\mu(\gamma)\vert 0\rangle$, where
$\Phi^\mu(\gamma)$ is the primary field. It is clear that it is
nothing but the zero mode of $\Phi^\mu$ in our cylindrical coordinates
$(\sif ,\sis )$. On general grounds, it immediately comes to mind to
introduce a new string vertex operator ${\cal O}^\mu_n=\oint_{C_0} dz\,
\ep^{n(i\sif -\sis )}{\cal V}_k^{(n)}\ep^{i\mu\varphi (z)}$  that corresponds 
to the $n$th mode of $\Phi^\mu$. 

\subsection{CFT representations}

What we have done so far was just to learn how to build the stringy
representations for $\su$ given by \eqref{su2}. Now we go over to
conformal field theory. In doing so, we must bear in mind that the
primary conformal fields may generally have dimensions that are
different from one, and then their integrals over $dz$ are not
invariant\footnote{Recall that we consider only the holomorphic
  sector.}. Let $\Phi(0)$ be the primary field setting at $z=0$. In
the framework of CFT (see \cite{BPZ}), the action of the Virasoro
(Kac-Moody) generators on $\Phi(0)$ is defined by
\begin{equation}
  L_n\bigl(\Phi(0)\bigr)=\oint_{C_0}dz\,z^{n+1}T(z)\Phi(0)
  \quad.
\end{equation}
Here $T(z)$ means the stress tensor of a world-sheet theory.

Likewise, we define the action for the stringy generators
\begin{equation}\label{cft-act}
  T\bigl(\Phi(0)\bigr)=\oint_{C_0}dz\,J(z)\Phi(0)
  \quad,\quad
  T=\oint_{C_0}dz\,J(z)
  \quad.
\end{equation}

Looking for highest weight vectors of $\su$, we see again that there
is a natural possibility:
\begin{equation}\label{hwv-cft2}
  \Phi^\mu(0)=V_q\ep^{ix\varphi(0)}
  \quad,
\end{equation}
where $V_q(0)$ should be interpreted as the exact vacuum operator of
the $(\beta,\gamma)$ system whose Bose sea level is $q$. Its OPE's
with $\beta$ and $\gamma$
\begin{equation*}
  \beta(z)V_q(0)\sim z^{q}
  \quad,\quad
  \gamma(z)V_q(0)\sim z^{-q}
  \quad,\quad
  z\rightarrow 0
\end{equation*}
result in $V_q(0)=\ep^{-q\sis (0)}$ \cite{FMS}.

Now we will briefly discuss the application of these ideas to the
construction of the unitary representations. The parameters $q$ and
$x$ are found by evaluating $\mathbf{H}_0\bigl(\Phi^\mu(0)\bigr)$ and
$\mathbf{k}\bigl(\Phi^\mu(0)\bigr)$. Eq.~\eqref{hwv-cft2} thus takes
the form
\begin{equation}\label{hwv-c2}
  \Phi^\mu(0)=\ep^{k\sis +i\mu\varphi(0)}
  \quad.
\end{equation}
Moreover, it can be shown that $\Phi^\mu$ turns out to be the highest
weight vector only if $k$ and $\mu$ obey \eqref{su2-u}.

In addition to the highest weight vector, there are other states that
may be easily determined. One state is
$\bigl(\mathbf{E}^-_0\bigr)^{\sqrt 2\mu }\Phi^\mu(0)\sim \ep^{k\sis
  -i\mu\varphi(0)}$, which corresponds to the lowest weight vector of
the $su(2)$ multiplet. The second is
$\bigl(\mathbf{E}^-_0\bigr)^{\sqrt 2\mu +1}\Phi^\mu(0)=0$, which
corresponds to the singular vector of the Verma module $V_\mu$
generated by the highest weight vector $\Phi^\mu(0)$. Finally, we have
$\bigl(\mathbf{E}^+_{-1}\bigr)^{k-\sqrt 2\mu+1}\Phi^\mu(0)=0$, which
corresponds to the second singular vector of $V_\mu$.  At this point a 
remark is in order.

It is well-known that the Fock space of $(\sif ,\sis )$ (large space)
is twice as large as the Fock space of $(\beta ,\gamma)$ (small space)
\cite{FMS}. To be more precise, it is defined as ${\cal F}(\beta
,\gamma)= \{v\in{\cal F}(\sif ,\sis ),\,\, \eta_0v=0\}$, where
$\eta_0=\oint_{C_0}dz\,\ep^{i\sif (z)}$. Our definition of the
space-time generators includes negative powers of $\gamma$ that
automatically leads to $\eta_0v\not=0$. Indeed,
$\eta_0\bigl(\gamma^n(0)\bigr)= \oint_{C_0}dz\,z^n\ep^{i\sif
  (z)+n(i\sif -\sis )(0)}$ that does not vanish for $n<0$. Thus we
have the large space\footnote{The situation is reminiscent of
  \cite{AF} where nontrivial modules are defined by negative powers of
  operators.}.

Now, what is special about the examples we have given in this section? To 
understand that, note that the same arguments that show that $\Phi^\mu(0)$ 
given by eq.~\eqref{hwv-c2} is the highest weight vector show that it is 
invariant under 
$\Phi^\mu(0)\rightarrow\ep^{\Lambda (i\sif -\sis )}\Phi^\mu(0)$ with an 
arbitrary parameter $\Lambda$. So, the highest weight vector takes the 
following form
\begin{equation}\label{v-c2}
  \Phi^\mu(0)=\ep^{i\Lambda\sif +(k-\Lambda)\sis +i\mu\varphi(0)}
  \quad.
\end{equation}
We considered two special cases: $\Lambda=\oh(k+\frac{1}{k}(2-k-\mu^2))$ 
and $\Lambda=0$.


\section{ Simply-laced algebras}
\renewcommand{\theequation}{3.\arabic{equation}}
\setcounter{equation}{0}

An obvious generalization of what we have just done is the following.
Consider a set of scalar fields $\{\varphi^i\}$ without background
charges.  This ensures that it is possible to construct the level $1$
representation of $\hat\mathfrak{g}$ whenever $\mathfrak{g}$ is a 
simply-laced algebra
i.e., it has equal length roots and hence is of the A, D, E type. This
is the so-called vertex operator construction \cite{Hal, vertex}. In
the Cartan-Weyl basis of $\mathfrak{g}$, the corresponding holomorphic currents
(generation functions) are represented by
\begin{equation}
  H^i(z)=i\pd\varphi^i (z)
  \quad,\quad
  E^{\alpha}(z)=\ep^{i\alpha\cdot\varphi (z)}c_\alpha 
  \quad,
\end{equation}
where $i=1,\dots ,\text{rank}\,\mathfrak{g}$, $\alpha$ is a root of 
$\mathfrak{g}$, $c_\alpha$ is a cocycle factor (see e.g., \cite{GO} for 
a review). All
roots have length $\sqrt 2$. The two-point functions of the free
fields are normalized as $\langle\varphi^i(z_1)\varphi^j(z_2)\rangle
=-\delta^{ij}\ln\zo$.

We take the space-time generators to be
\begin{equation}\label{g}
  \mathbf{H}^i_n=\oint_{C_0} dz\,\gamma^n H^i(z)
  \quad,\quad 
  \mathbf{E}^\alpha_n=\oint_{C_0} dz\,\gamma^n E^{\alpha}(z)
  \quad,\quad 
  n\in\mathbf{Z}\quad,
\end{equation}
which is a natural generalization to arbitrary $\mathfrak{g}$ of what we had 
for $su(2)$. It is clear that for $\gamma=const$, the definition agrees
with the definition of the generators for a centerless affine algebra
\cite{Kac}.

Now, let us calculate the commutation relations of the space-time
generators.  Using conformal techniques, we obtain
\begin{equation}\label{g-st}
  \begin{split}
    [\mathbf{H}^i_n,\mathbf{H}^j_m]&
    =\mathbf{k}n\delta^{ij}\delta_{n+m,0} \quad,\quad
    [\mathbf{H}^i_n,\mathbf{E}^{\alpha}_m]=\alpha^i\mathbf{E}^{\alpha}_{m+n}
    \quad, \\
    [\mathbf{E}^{\alpha}_n,\mathbf{E}^{\beta}_m]& =
    \begin{cases}
      \varepsilon(\alpha,\beta)\mathbf{E}^{\alpha+\beta}_{n+m} &
      \text{ if
        $\alpha+\beta$ is a root of $\mathfrak{g}$}\quad, \\
      \mathbf{k}n\delta_{n+m,0}+\alpha\cdot\mathbf{H}_{n+m} &
      \text{ if $\alpha+\beta=0$}\quad,\\
      0 & \text{ otherwise}\quad,
    \end{cases} 
  \end{split}
\end{equation}
together with
\begin{equation}
  [\mathbf{H}^i_n,\mathbf{k}]=[\mathbf{E}^{\alpha}_n,\mathbf{k}]=0\quad.
\end{equation}
Here $\varepsilon(\alpha,\beta)$ equals $\pm 1$ and satisfies certain
consistency conditions (see e.g., \cite{GO}).

Before continuing our discussion, we will make a detour and recall
some basics facts on the unitary hight weight representations of 
$\hat\mathfrak{g}$. It is well-known that $\hat\mathfrak{g}$ possesses 
the unitary highest weight representations that are defined as follows. Let
$\vert\mu\rangle$ be the highest weight vector satisfying
\begin{equation}\label{hwv-g}
  \mathbf{H}^i_0\vert\mu\rangle=\mu^i\vert\mu\rangle
  \quad,\quad
  \mathbf{k}\vert\mu\rangle=k\vert\mu\rangle
  \quad,\quad
  \mathbf{E}^\alpha_0\vert\mu\rangle=0
  \quad\text{for}\quad \alpha>0
  \quad,\quad
  \mathbf{H}^i_n\vert\mu\rangle=\mathbf{E}^{\alpha}_n\vert\mu\rangle=0
  \quad\text{for}\quad n>0
  \quad.
\end{equation}
The representation $V_\mu$ is built as the Verma module over $\hat\mathfrak{g}$
generated by the vector $\vert\mu\rangle$. There is a simple set of
necessary and sufficient conditions for there to be a unitary highest weight 
representation of $\hat\mathfrak{g}$ in which $k$ and $\mu$ take special 
values; it is
\begin{equation}\label{g-u}
  k\in\mathbf{N}\quad,\quad 0\leq\alpha\cdot\mu\leq k\quad\text{with}\quad 
  \alpha\cdot\mu\in\mathbf{N}
  \quad.
\end{equation}

It is straightforward to rewrite the space-time generators of 
$\hat\mathfrak{g}$ in terms of the scalar fields. Using \eqref{bos}, we find
\begin{equation}\label{g-b}
  \mathbf{H}^i_n=\oint_{C_0} dz\,\ep^{n(i\sif -\sis )}H^i(z)
  \quad,\quad 
  \mathbf{E}^\alpha_n=\oint_{C_0} dz\,\ep^{n(i\sif -\sis )}E^{\alpha}(z)
  \quad.
\end{equation}

\subsection{Stringy representations}

First, to complete our discussion of the space-time generators
\eqref{g}, we should perhaps point out that in the problem at hand,
they obey
\begin{equation}\label{phys-g}
  [L_n,\mathbf{H}^i_m]=[L_n,\mathbf{E}^\alpha_m]=0
  \quad,
\end{equation}
where $L_n$ are the Virasoro generators on the world-sheet i.e.,
\begin{equation}\label{vir-g}
  L_n=\oint_{C_0} dz\,z^{n+1}\bigl(\beta\pd\gamma-
  \oh\pd\varphi\cdot\pd\varphi\bigr)(z)
  \quad.
\end{equation}
Thus, they are the string vertex operators.

Now, our task is to build the unitary highest weight representations
for $\hat\mathfrak{g}$ given by \eqref{g-b}. In the $SU(2)$ case that we have
considered, it is not hard to guess how to do this. In a similar way
we can try to define ${\cal O}^\mu$ as
\begin{equation}\label{g-hwv}
  {\cal O}^\mu=\oint_{C_0} dz\,\ep^{ix^+\sif +x^-\sis }\,
  \ep^{ix\cdot\varphi (z)}
  \quad.
\end{equation}
The only difference is therefore that $x$ has become a
$\text{rank}\,\mathfrak{g}$-dimensional vector.

The next step in finding the highest weight vector is to adopt the
prescription \eqref{act} in the present situation. By using this definition, 
the parameters $x^{\pm}$, $x^i$ may be explicitly determined as
follows. According to \eqref{hwv-g}, the action $\mathbf{H}^i_0$ on
${\cal O}^\mu$ is given by $\mu^i{\cal O}^\mu$, hence $x^i=\mu^i$.
Similarly, $\mathbf{k}$ acts as $k{\cal O}^\mu$ that results in 
$x^++x^-=k$.  Moreover, we know that ${\cal O}^\mu$ must commute with
${\cal L}_n$ given by \eqref{vir-g}. The last is true only for
$x^{\pm}$ such that $\mu\cdot\mu+k(x^+-x^-+1)=2$. Thus, the allowed
values of $x^{\pm}$ are $\oh(k\pm\frac{1}{k}(2-k-\mu\cdot\mu))$.
Before going on, let us note that it is also straightforward to write
the action in the form of \eqref{act1} by rescaling ${\cal
  O}^\mu\rightarrow {\cal O}^\mu c_\mu$, where $c_\mu$ is a cocycle
factor. In this case, it is given by $c_\mu=(-)^{\mu\cdot p+x^+
  \rho^1-x^2\rho^2}(-)^{\mu *p}$ with
$p^i=i\oint_{C_0}dz\,\pd\varphi^i(z)\,,\,\,
\rho^1=i\oint_{C_0}dz\,\pd\sif
(z)\,,\,\,\rho^2=-\oint_{C_0}dz\,\pd\sis (z)$.  The product $*$ is
defined as $\mu *\alpha=\sum_{i>j}n_im_j\boldsymbol{\mu}^i
\boldsymbol{\alpha}^j$, where $\mu=\sum_i n_i\boldsymbol{\mu}^i$,
$\alpha= \sum_jm_j\boldsymbol{\alpha}^j$; $\boldsymbol{\mu}^i$,
$\boldsymbol{\alpha}^j$ are the fundamental weights and simple roots
of $\mathfrak g$, respectively.

Let us now examine the operators ${\cal O}^\mu$ more closely. The
basic tool is conformal field technique. The evaluation of the actions
for $\mathbf{H}^i_n$ gives
\begin{equation}\label{act-H}
  \mathbf{H}^i_n\bigl({\cal O}^\mu\bigr)=\oint_{C_0}dz_2
  \oint_{C_{z_2}}dz_1\,(\zo )^{nk-1}\Bigl(\mu^i+O(\zo )\Bigr)
  \ep^{n(i\sif -\sis )(z_1)}{\cal V}_k\ep^{i\mu\cdot\varphi (z_2)}
  \quad.
\end{equation}
It immediately follows that the contour integrals are well-defined
only for $k\in\mathbf{Z}$. This is in harmony with what we have, since
our topological construction of $k$ allows only integer values.
Moreover, we see that $\mathbf{H}^i_n$ with $n>0$ annihilate ${\cal
  O}^\mu$ because the integrand is non-singular for $k>0$.  Otherwise,
$\mathbf{H}^i_n\bigl({\cal O}^\mu\bigr)$ would not vanish.  In
particular, $\mathbf{H}^i_0$ act as claimed before.

In a similar way one can also find the actions for
$\mathbf{E}^\alpha_n$.  These are given by
\begin{equation}\label{act-E}
  \mathbf{E}^{\alpha}_n\bigl({\cal O}^\mu\bigr)=\oint_{C_0}dz_2
  \oint_{C_{z_2}}dz_1\,(\zo )^{nk+\alpha\cdot\mu}
  \ep^{i\alpha\cdot\varphi}\ep^{n(i\sif -\sis )(z_1)}
  {\cal V}_k\ep^{i\mu\cdot\varphi (z_2)}
  \quad.
\end{equation}
This time the contour integrals are well-defined if $\alpha\cdot\mu\in
\mathbf{Z}$. By considering $\alpha >0$ and $n>0$, we see that the
corresponding $\mathbf{E}^\alpha_n$ annihilate ${\cal O}^\mu$ for
$\alpha\cdot\mu\geq 0$. On the other hand, we see that there is a
restriction $k\geq \alpha\cdot\mu$ that guarantees the vanishing of
all $\mathbf{E}^{\alpha}_n\bigl({\cal O}^\mu\bigr)$ for $n>0$ and
negative roots.  Thus we have recovered the standard conditions
\eqref{g-u} and have succeeded in finding the highest weight vectors
for the unitary representations of $\hat\mathfrak{g}$.

It remains to show that the Verma modules generated by such vectors
contain only the string vertex operators. The analysis proceeds
exactly as in the previous section. The basis point is that all
operators $\mathbf{H}^i_n$, $\mathbf{E}^\alpha_n$ and ${\cal O}^\mu$
are stringy. As a consequence, any state from the Verma module turns
out to be stringy too. This completes our construction of the unitary
representations for $\hat\mathfrak{g}$.

In addition to the highest weight vectors built above, there are other 
important vectors that may be easily determined. Let us give a couple of 
examples. For instance, we can find the lowest weight vectors for the 
$\mathfrak{g}$ multiplet. A short computation shows that these are 
$\oint_{C_0} dz\,{\cal V}_k\ep^{-i\mu\cdot\varphi (z)}$. Moreover, having 
the set of roots $\{\alpha\}$, we can choose a simple root 
$\boldsymbol{\alpha}$ among them. It is well-known that the singular vector 
associated with this simple root is 
$\bigl(\mathbf{E}^{-\boldsymbol{\alpha}}_0\bigr)
^{\boldsymbol{\alpha}\cdot\mu+1 }\vert\mu\rangle$. A simple dimensional 
analysis shows that $\bigl(\mathbf{E}^{-\boldsymbol{\alpha}}_0\bigr)
^{\boldsymbol{\alpha}\cdot\mu+1 }\bigl({\cal O}^\mu\bigr)=0$ within our 
construction. 

\subsection{CFT representations}

We will briefly summarize the generalization for an arbitrary
simply-laced algebra $\mathfrak g$.

First of all, the highest weight vector is given by
\begin{equation}\label{hwv-cft-g}
  \Phi^\mu(0)=V_q\ep^{ix\cdot\varphi(0)}
  \quad,
\end{equation}
where $V_q(0)$ is again the $q$-vacuum operator of the
$(\beta,\gamma)$ system. The primary field is setting at $z=0$. The
parameters $q$ and $x$ are found by evaluating the $\mathbf{H}^i_0$
and $\mathbf{k}$ on \eqref{hwv-cft-g}. Thus, the hight weight vector
becomes
\begin{equation}\label{hwv-c-g}
  \Phi^\mu(0)=\ep^{k\sis +i\mu\cdot\varphi(0)}
  \quad.
\end{equation}

Moreover, a simple analysis again shows that $\mu^i$ and $k$ must obey
\eqref{g-u}, i.e. the representation we built is unitary.

In addition to the highest weight vector, the lowest weight vector of the 
$\mathfrak g$ multiplet may be easily determined. It is given by
$\Phi^{-\mu}(0)\sim\ep^{k\sis -i\mu\cdot\varphi(0)}$. Moreover, the singular 
vectors $\bigl(\mathbf{E}^{-\boldsymbol{\alpha}}_0\bigr)
^{\boldsymbol{\alpha}\cdot\mu+1 }\Phi^\mu(0)$ vanish.


\section{The N=4 case}
\renewcommand{\theequation}{4.\arabic{equation}}
\setcounter{equation}{0} We now want to reconsider the above
discussion in the context of supersymmetry. One of the most
interesting problems of this kind concerns the $N=4$ superconformal
algebra. There are at least two reasons to say so. The first is its
appearance in a description of superstring propagation on $\ads$
\cite{Mal}. The second is that the algebra includes $\su$ as a
subalgebra, so we need a slight generalization of what we have done in
section 2. First let us make a short detour to summarize several
results obtained in \cite{A}.

According to the arguments by Ito and further elucidated by ourselves,
adding fermionic coordinates geometrically turns space-time into a
superspace. More precisely, this means that we add to the
$(\beta,\gamma)$ system a set of first order fermionic systems of
weights $(1,0)$. To get the space-time $N=4$ superconformal algebra,
it is enough to add two systems, say $(\eta,\xi)$ and $(\te ,\txi )$.
According to \cite{FMS}, any first order fermionic system can be
bosonized by a scalar field with a proper background charge. In
particular, the fermionic fields in the problem at hand are given by
\begin{equation*}
  \xi(z)=\ep^{-i\phi (z)}
  \quad,\quad
  \eta(z)=\ep^{i\phi (z)}
  \quad,\quad
  \txi (z)=\ep^{-i\tilde\phi (z)}
  \quad,\quad
  \te (z)=\ep^{i\tilde\phi (z)}
  \quad.
\end{equation*}
It is advantageous to introduce new fields as
\begin{equation*}
  \vaf (z)=\frac{1}{\sqrt 2}(-\phi+\tilde\phi)(z)
  \quad,\quad
  \vas (z)=\frac{1}{\sqrt 2}(\phi+\tilde\phi)(z)
  \quad,
\end{equation*}
which leads to the world-sheet stress-tensor $T(z)=-\oh\pd\vaf\pd\vaf-
\oh\pd\vas\pd\vas-\frac{i}{\sqrt 2}\pd^2\vas (z)$. Thus we have the
$c=1$ theory and an additional theory with $c=-5$.  To generalization
our treatment in section 2, we therefore need to add the scalar field
$\vas$ with the background charge. Now it is obvious how to build the
$\su$ subalgebra of the space-time $N=4$ algebra. It is simply given
by \eqref{su2}. As to the other generators, they can be found in
\cite{A}.  In terms of the scalar fields, these are\footnote{Our
  conventions for the bosonic generators of $N=4$ will be those of
  \cite{A}. In particular, $\mathbf{H}_n\rightarrow \sqrt
  2\mathbf{T}^0_n$ and
  $\mathbf{E}^{\pm}_n\rightarrow\mathbf{T}^{\pm}_n$.}
\begin{equation}\label{qn4}
  \begin{split}
    \mathbf{L}_n &=\oint_{C_0}dz\,\ep^{n(i\sif -\sis )}\Bigl(-\pd\sis
    + \frac{i}{\sqrt 2}(n+1)\pd\vas\Bigr)(z)
    \quad,\quad \\
    \mathbf{T}_n^0 &=\frac{i}{\sqrt 2}\oint_{C_0}dz\,\ep^{n(i\sif
      -\sis )} \pd\vaf (z) \quad,\quad
    \mathbf{T}_n^{\pm}=\oint_{C_0}dz\,\ep^{n(i\sif -\sis )} \ep^{\pm
      i\sqrt 2\vaf (z)}
    \quad,\\
    \mathbf{G}^{\pm}_r &=\pm i\sqrt2\oint_{C_0}dz\,\ep^{(r+\oh )(i\sif
      -\sis )} \ep^{\frac{i}{\sqrt 2}(\pm\vaf +\vas
      )(z)}\ep^{\frac{i\pi}{\sqrt 2}p^1}
    \quad,\quad \\
    \bar\mathbf{G}^{\pm}_r& =-\sqrt2\oint_{C_0}dz\,\ep^{(r-\oh )(i\sif
      -\sis )} \ep^{\frac{i}{\sqrt 2}(\pm\vaf -\vas
      )}\Bigl(\pd\sis\mp\frac{i}{\sqrt 2} (r-\oh )\pd\vaf
    -\frac{i}{\sqrt 2}(r+\frac{3}{2} )\pd\vas\Bigr)(z)
    \ep^{-\frac{i\pi}{\sqrt 2}p^1} \quad.
  \end{split}
\end{equation}
Here we explicitly write the cocycle factors in the last two
expressions.

The corresponding commutation relations are
\begin{equation}\label{N4}
  \begin{split}
    [\mathbf{L}_n,\mathbf{L}_m]& =(n-m)\mathbf{L}_{n+m}+
    \frac{\mathbf{c}}{12}(n^3-n)\delta_{n+m,0}
    \quad,\\
    [\mathbf{T}_n^a,\mathbf{T}_m^b]&
    =\frac{\mathbf{c}}{12}ng^{ab}\delta_{n+m,0}
    +f^{ab}_c\mathbf{T}^c_{n+m}
    \quad,\\
    \{\mathbf{G}^\alpha_r,\bar\mathbf{G}_s^\beta\}&=
    \frac{\mathbf{c}}{3} (r^2-\oq )\eta^{\alpha\beta} \delta_{r+s,0}+
    2\eta^{\alpha\beta}\mathbf{L}_{r+s}+
    4(r-s)(\sigma^a)^{\alpha\beta}\mathbf{T}_{r+s}^bg_{ab}
    \quad,\\
    [\mathbf{L}_n,\mathbf{T}_m^a]& =-m\mathbf{T}_{n+m}^a \quad,\quad
    [\mathbf{L}_n,\mathbf{G}_r^\alpha]=(\frac{n}{2}-r)\mathbf{G}^\alpha_{r+n}
    \quad,\quad [\mathbf{L}_n,\bar\mathbf{G}_r^\alpha]=
    (\frac{n}{2}-r)\bar\mathbf{G}^\alpha_{r+n}
    \quad,\quad \\
    [\mathbf{T}_n^a,\mathbf{G}^\alpha_r]& =
    (\sigma^a)^\alpha_\beta\mathbf{G}^\beta_{r+n} \quad,\quad
    [\mathbf{T}_n^a,\bar\mathbf{G}^\alpha_r]=
    (\bar\sigma^a)^\alpha_\beta\bar\mathbf{G}^\beta_{r+n} \quad,
  \end{split}
\end{equation}
where
$g^{00}=1,\,\,g^{+-}=g^{-+}=2;\,\,f^{0+}_+=f^{-0}_-=1,\,\,f^{+-}_0=2;\,\,
a,b,c=0,\pm$; $\eta^{+-}=\eta^{-+}=1$, $\alpha,\,\beta=\pm$;
$(\sigma^a)^{\alpha\beta}=(\sigma^a)^{\alpha}_{\gamma}\eta^{\gamma\beta}$;
and the tensors $(\sigma^a)^\alpha_\beta$ are in the representation
\begin{equation*}
  (\bar\sigma^+)^-_+=(\bar\sigma^-)^+_-=-(\sigma^+)^-_+=-(\sigma^-)^+_-=1
  \quad,\quad
  (\bar\sigma^0)^+_+=-(\bar\sigma^0)^-_-=(\sigma^0)^+_+=-(\sigma^0)^-_-=\oh
  \quad.
\end{equation*}
Moreover,
\begin{equation*}
  [\mathbf{L}_n,\mathbf{k}]=[\mathbf{T}^a_n,\mathbf{k}]=
  [\mathbf{G}_r^\alpha,\mathbf{k}]=[\bar\mathbf{G}_r^\alpha,\mathbf{k}]=0
  \quad,\quad
  \mathbf{c}=6\mathbf{k}
  \quad,
\end{equation*}
where $\mathbf{k}$ is given by \eqref{kb}.

At this point, it is necessary to make a remark. One of the important
statements about the $N=4$ superconformal algebra was the following
observation by Schwimmer and Seiberg \cite{SS}. There is an infinite
set of independent sectors (algebras) of $N=4$ labeled by an angular
valued parameter corresponding to the conjugate classes of $SU(2)$.
Thus representations for a given value of the central charge $c$ are
not equivalent for different values of the parameter. In our
discussion, we restrict ourselves to the Neveu-Schwarz sector, so
$n,m\in\mathbf{Z};\,\,r,s\in\mathbf{Z}+\oh$ in \eqref{N4} and below.
Of course, it seems interesting to develop our approach for a general
case elsewhere.

It is well-known (see \cite{ET}) that the $N=4$ superconformal algebra
possesses the unitary highest weight representations that are defined
as follows. Let $\vert h,j\rangle$ be the highest weight vector,
namely
\begin{equation}\label{N4hwv}
  \begin{split}
    \mathbf{L}_0\vert h,j\rangle &=h\vert h,j\rangle \quad,\quad
    \mathbf{T}^0_0\vert h,j\rangle =j\vert h,j\rangle \quad,\quad
    \mathbf{k}\vert h,j\rangle =k\vert h,j\rangle \quad,\quad
    \mathbf{T}^+_0\vert h,j\rangle =0
    \quad,\quad \\
    \mathbf{L}_n\vert h,j\rangle &=\mathbf{T}^a_n\vert h,j\rangle =0
    \quad,\quad\text{for}\quad n>0 \quad,\quad
    \mathbf{G}^\alpha_r\vert h,j\rangle =\bar\mathbf{G}^\alpha_r\vert
    h,j\rangle =0 \quad,\quad\text{for}\quad r>\oh \quad.
  \end{split}
\end{equation}
The representation $V_{h,j}$ is built as the Verma module over $N=4$
generated by the vector $\vert h,j\rangle$. It turns out that there
exist two classes of the unitary representations:
\begin{alignat}{3}
  \text{(A)}\,&\text{Massive representations.}& \quad h
  &>j\quad,&\quad j&=0,\oh ,\dots ,\oh k-\oh
  \quad. \label{mass} \\
  \text{(B)}\,&\text{Massless representations.}& \quad h&=j
  \quad,&\quad j&=0,\oh,\dots ,\oh k \quad.\label{chiral}
\end{alignat}

We pause here to point out a subtlety that we have not revealed so
far. The preceding analysis of the commutation relations and the
representations is valid only for well-defined integrands. In other
words, the integrands of \eqref{cft-act} for the CFT representations
(or \eqref{act} for the stringy representations) are single-valued as
$z\,(z_1)$ goes around the origin $(z_2)$ so that the integration
contour closes (see Fig.1).
%
\vspace{.4cm}
\begin{figure}[ht]
\begin{center}
\includegraphics{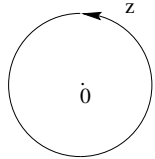}
\caption{The integration contour in eq.~\eqref{cft-act}.}
\label{fig:graph1}
\end{center}
\end{figure}
\vspace{-0.3cm} 

\noindent This is compatible with the claim that the level $k$ and the 
weights take special discrete values. It is amusing that these values 
correspond to unitary representations (see \eqref{g-u}). In other words, 
the local objects of the world-sheet theory result in the unitary
space-time theory.

On the other hand, suppose that we replace the discrete values of the
weights by generic ones. The main point is that the integration
contour no longer closes. A possible way to bring the situation under
control is to define a new integration contour as shown in Fig.2
%
\vspace{.4cm}
\begin{figure}[ht]
\begin{center}
\includegraphics{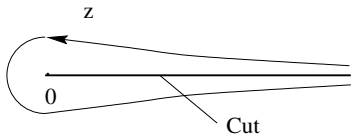}
\caption{The modified integration contour for eq.~\eqref{cft-act}.}
\label{fig:graph2}
\end{center}
\end{figure}
\vspace{-0.3cm} 

\noindent The object just defined is similar to the screened vertex operator 
of CFT, but one detail is different, namely we have a representation
of $\mathfrak{g}$ rather than $U_q(\mathfrak{g})$, as is usual in CFT.
Note that such construction is nonlocal since it defines the bilocal
objects that depend on two points $0$ and $\infty$, where the cut
ends. The last does not sound reasonable within string theory. For the
unitary representations of simply-laced algebras, this is not so
because they have the discrete values of the weights. However, the
massive representations of $N=4$ exhibit a continuum in the spectrum
of conformal dimensions that can affect the locality of a theory on
the world-sheet\footnote{For a recent attempt to discuss a continuum
  in the spectrum of conformal dimensions within the D1/D5 system (in
  particular, strings on $\ads$), see \cite{SW}.}. At the present time
it is not known whether the situation may be taken under control. So
we are bound to learn something if we succeed.

\subsection{CFT representations}

Now we want to build the highest weight representations on the $N=4$
algebra within CFT. In this setting, it is natural to start by writing
down the highest weight vector in the following form:
\begin{equation}\label{hwv-cft-N4}
  \Phi^{h.j}(0)=V_q\ep^{ix\cdot\varphi(0)}
  \quad.
\end{equation}
Note that the approach we use is similar to that of subsection 3.2,
but the problem is different. In particular, we have the scalar field
$\vas$ with the background charge.

The generators $\mathbf{k}$, $\mathbf{L}_0$ and $\mathbf{T}^0_0$ must
act on $\Phi^{h.j}$ according to \eqref{N4hwv}. For this to be so, the
Bose sea level $q$ must have the value $-k$. Also, the allowed values
of $x^1$, $x^2$ must be $\sqrt 2j$ and $\sqrt 2(h-k)$, respectively.
Thus, the highest weight vector is
\begin{equation}\label{hwv-N4}
  \Phi^{h.j}(0)=\ep^{k\sis +i\sqrt 2j\vaf +i\sqrt 2(h-k)\vas (0)}
  \quad.
\end{equation}
To be more precise about this, to convince the reader that this is
indeed the case, we have to prove the statement. This is proved by
explicitly evaluating actions of the $N=4$ generators on the highest
weight vector.

For $\mathbf{L}_n$, we get
\begin{equation}\label{4-L}
  \mathbf{L}_n\bigl(\Phi^{h.j}(0)\bigr)=\oint_{C_0}dz\,
  z^{nk-1}\Bigl( h+(h-k)n+O(z)\Bigr)
  \ep^{in\sif +(k-n)\sis +i\sqrt 2j\vaf +i\sqrt 2(h-k)\vas (0)}
  \quad.
\end{equation}
It immediately follows that the contour integral is well-defined only
for $k\in\mathbf{Z}$. Moreover, the $\mathbf{L}_n$ with $n>0$
annihilate $\Phi^{h.j}$ only if $k>0$. That is in harmony with what we
had before.

Now, let us see what happens when $\mathbf{T}^a_n$ act on the highest
weight vector. For $\mathbf{T}^{\pm}_n$, the actions are given by
\begin{equation}\label{4T+-}
  \mathbf{T}^{\pm}_n\bigl(\Phi^{h.j}(0)\bigr)=\oint_{C_0}dz\,
  z^{nk\pm 2j}\Bigl(1+O(z)\Bigr)
  \ep^{in\sif +(k-n)\sis +i\sqrt 2(j\pm 1)\vaf +i\sqrt 2(h-k)\vas (0)}
  \quad.
\end{equation}
This means that the contour integral is well-defined only for
$2j\in\mathbf{Z}$. Moreover, one can specify $n$ to determine the
allowed values of $j$. For $\mathbf{T}^+_0$,
$\mathbf{T}^+_0\bigl(\Phi^{h.j}(0)\bigr)=0$, which implies
$2j\in\mathbf{N}$.  For $\mathbf{T}^-_1$,
$\mathbf{T}^-_1\bigl(\Phi^{h.j}(0)\bigr)$ must also vanish, which
leads to $j\leq\frac{k}{2}$. The latter is exactly the property of the
massless representations \eqref{chiral}. For $\mathbf{T}^0_n$, we have
\begin{equation}\label{4T0}
  \mathbf{T}^0_n\bigl(\Phi^{h.j}(0)\bigr)=\oint_{C_0}dz\,
  z^{nk-1}\Bigl(j+O(z)\Bigr)
  \ep^{in\sif +(k-n)\sis +i\sqrt 2j\vaf +i\sqrt 2(h-k)\vas (0)}
  \quad.
\end{equation}
We see that all of the integrands are non-singular for $n>0$, so 
such the $\mathbf{T}^0_n$ annihilate $\Phi^{h.j}(0)$. As to
$\mathbf{T}^0_0$, it acts as claimed before.

Now, let us turn to the fermionic generators. In this discussion, we
neglect the cocycle factors as they result in irrelevant numerical
factors. For the fermionic generators $\mathbf{G}^{\pm}_r$, we have
\begin{equation}\label{4G}
  \mathbf{G}^{\pm}_r\bigl(\Phi^{h.j}(0)\bigr)=\oint_{C_0}dz\,
  z^{k(r-\oh )+h\pm j}\Bigl(1+O(z)\Bigr)
  \ep^{i(r+\oh )\sif +(k-r-\oh )\sis +i\sqrt 2(j\pm\oh )\vaf 
+i\sqrt 2(h-k+\oh )\vas (0)}
  \quad.
\end{equation}
We see that the integrand is single-valued i.e., the integration
contour closes, if $h\pm j\in\mathbf{Z}$. This requirement and the
requirement $j\leq\frac{k}{2}$ found in above imply, together, that we
only get the massless representations within our construction. In a
similar way, for the $\bar\mathbf{G}^{\pm}_r$, we have
\begin{equation}\label{4bG}
  \begin{split}
    \bar\mathbf{G}^{\pm}_r\bigl(\Phi^{h.j}(0)\bigr)&=\oint_{C_0}dz\,
    z^{k(r+\oh )-h\pm j-1}\Bigl(k\pm j(r-\oh )+(r+\frac{3}{2})(h-k)
    +O(z)\Bigr)\\
    &\times\ep^{i(r-\oh )\sif +(k-r+\oh )\sis +i\sqrt 2(j\pm\oh )\vaf
      + i\sqrt 2(h-k-\oh )\vas (0)} \quad,
  \end{split}
\end{equation}
that gives rise to $h\pm j\in\mathbf{Z}$. As the result, we are again 
restricted to the massless representations.

Thus, as promised in the introduction to this section, the local
objects of the world-sheet theory result in the unitary space-time
theory; in contrast to the affine Lie algebras (at least A,D,E type),
however, we have not found the all unitary representations. The
problem occurs with the massive representations when there exists a
continuum in the spectrum of conformal dimensions.

To understand what is special about these massless representations, it
helps to consider their singular vectors. First, we consider the
``bosonic'' singular vectors i.e., the ones obtained by applying
bosonic generators to the highest weight vector. These are simply
$\bigl(\mathbf{T}^-_0\bigr)^{2j+1}\bigl(\Phi^{j.j}(0)\bigr)$ and
$\bigl(\mathbf{T}^+_0\bigr)^{k-2j+1}\bigl(\Phi^{j.j}(0)\bigr)$. The
analysis proceeds exactly as in subsection 2.1 and shows that the
vectors vanish.  Moreover, being the massless representation, the
Verma module generated by $\Phi^{j.j}(0)$ possesses the ``fermionic''
singular vectors \cite{ET1}. The vectors of interest are
$\bar\mathbf{G}^+_{-\oh }\bigl(\Phi^{j.j}(0)\bigr)$ and
$\mathbf{G}^+_{-\oh }\bigl(\Phi^{j.j}(0)\bigr)$. Setting $r=-\oh$ as
well as $h=j$ in \eqref{4bG}, we see that the integrand is regular. As
a result, we get $\bar\mathbf{G}^+_{-\oh
  }\bigl(\Phi^{j.j}(0)\bigr)=0$. On the other hand, equation
\eqref{4G} gives $\mathbf{G}^+_{-\oh
  }\bigl(\Phi^{j.j}(0)\bigr)\not=0$. Thus, the representation we built
is reducible.

Of course, one can try to perform the same analysis for the stringy
representations. However, we do not see much sense in this because the
construction does not catch the massive representations. So, something
is missing in our understanding.

\subsection{More comments on the underlying world-sheet algebra}

Finally, we will present a description of the underlying world-sheet
algebra that is more precise than we gave in \cite{A}. To begin with,
we review the original analysis. In the framework of the free fields
$(\beta,\gamma)$, $(\eta,\xi)$, $(\te ,\txi )$, there exists the
following representation of the affine superalgebra $\ssl$ at the
level $k=1\,$\footnote{Our conventions will be those of \cite{A} (see
  eq.(3.28)).}
\begin{alignat}{4}\label{sl2/1}
  J^+(z) & =\xi\te (z)\,, & \quad J^-(z) & =\txi\eta (z)\,, & \quad
  J^0(z) & =\oh (\xi\eta -\txi\te )(z)\,, & \quad J^3(z) & =
  \beta\gamma+\oh (\xi\eta +\txi\te )(z)\,,\notag
  \\
  j^+(z) & = \frac{1}{\sqrt 2}\gamma\te (z)\,, & \quad j^{\pp }(z) & =
  \frac{1}{\sqrt 2}\xi\beta (z)\,, & \quad j^-(z) & = \frac{1}{\sqrt
    2}\txi \beta\,, & \quad j^{\mm}(z) & =\frac{1}{\sqrt 2}\gamma\eta
  (z) \quad,
\end{alignat}
whose OP expansions are
\begin{align}
  J^a(z_1)J^b(z_2)&=\frac{1}{\zt }\frac{g^{ab}}{2}+
  \frac{1}{\zo}f^{ab}_cJ^c(z_2)+O(1)\quad,\notag \\
  J^a(z_1)j^\alpha(z_2)&=\frac{1}{\zo}(\sigma^a)^\alpha_\beta
  j^\beta(z_2)+
  O(1)\quad, \\
  j^\alpha(z_1)j^\beta(z_2)& =\frac{1}{\zt
    }\frac{\varepsilon^{\alpha\beta}}{2}+
  \frac{1}{\zo}(\sigma^a)^{\alpha\beta}J_a(z_2)+ O(1)\quad, \notag
\end{align}  
where
$g^{00}=1,\,\,g^{+-}=g^{-+}=2,\,\,g^{33}=-1;\,\,f^{0+}_+=f^{-0}_-=1,\,\,
f^{+-}_0=2;\,\,a,b,c=0,\pm ,3$. $\varepsilon$ is the antisymmetric
tensor with $\varepsilon^{-+}=\varepsilon^{\pp\mm}=1$, where
$\alpha,\,\,\beta=\pm ,\,\dot\pm$. The tensors
$(\sigma^a)^\alpha_\beta$ are in the representation
\begin{alignat*}{5}
  (\sigma^0)^+_+&=(\sigma^0)^{\pp }_{\pp }&=-(\sigma^0)^-_-&=
  -(\sigma^0)^{\mm }_{\mm }&=\oh
  \quad, \\
  (\sigma^+)^-_{\pp}&=(\sigma^-)^{\pp }_-&=-(\sigma^+)^{\mm }_+&=
  -(\sigma^-)^+_{\mm }&=1
  \quad,\\
  (\sigma^3)^+_+&=(\sigma^3)^{\mm }_{\mm }&=-(\sigma^3)^{\pp }_{\pp
    }&= -(\sigma^3)^-_-&=\oh \quad.
\end{alignat*}
Moreover, $(\sigma^a)^{\alpha\beta}=(\sigma^a)^\alpha_\gamma
\varepsilon_{\gamma\beta}$.

Of course, the space-time $N=4$ generators can be rewritten in terms
of the world-sheet $\ssl$ generators, which provides a new relation
between these algebras \cite{A}. Although such construction looks in
many ways attractive, we have to stress a subtle point here. The
central charge of $\ssl$ identically equals zero while the free fields
have a total central charge $c_{\text t}=-2$. So, some world-sheet
degrees of freedom are missing. To understand what happened, it helps
to bosonize the fermionic fields as we have done in subsection 4.1.
Then the bosonic generators are rewritten as
\begin{equation}\label{bN4}
  J^0(z)=\frac{i}{\sqrt 2}\pd\vaf (z)
  \quad,\quad
  J^\pm(z)=\ep^{\pm i\sqrt 2\vaf (z)}
  \quad,\quad
  J^3(z)=\beta\gamma -\frac{i}{\sqrt 2}\pd\vas (z)
  \quad.
\end{equation}
Due to the above formula, we can think of $\vaf $ as the scalar field
that provides the world-sheet $\su$ algebra at the level $1$ (see
\eqref{su-2}).  As for $J^3$, it corresponds to the $U(1)$ current in
the Wakimoto free field representation of $\so$ \cite{Wak}. However,
the metric signature is now $(-,+,+)$. The rest of the currents is
given by
\begin{equation}\label{Wr}
  {\cal J}^-(z)=\beta (z)
  \quad,\quad
  {\cal J}^+(z)=\beta\gamma^2-i\sqrt 2\gamma\vas +\pd\gamma (z)
  \quad.
\end{equation}
It is not difficult to find the level of $\so$ which turns out to be
$1$.  Thus, we have the world-sheet $\so\times\su$ algebra at the
levels $1$. The central charge of this model exactly equals $-2$.
Moreover, the Sugawara construction results in the stress-tensors of
the free fields. This allows us to say that all degrees of freedom are
taken into account. Furthermore, the fermionic generators
\eqref{sl2/1} are simply given in terms of $\so\times\su$ primaries
by
\begin{equation}
  j^+(z)=\frac{1}{\sqrt 2}\phi^{\oh }V^{\oh}(z)
  \,\,\,,\,\,\,
  j^{\mm }(z)=\frac{1}{\sqrt 2}\phi^{\oh }\tilde V^{\oh}(z)
  \,\,\,,\,\,\,
  j^{\pp }(z)=\frac{1}{\sqrt 2}\tilde\phi^{\oh }V^{\oh}(z)
  \,\,\,,\,\,\,
  j^-(z)=\frac{1}{\sqrt 2}\tilde\phi^{\oh }\tilde V^{\oh}(z)
  \,\,\,.
\end{equation}
Here $V^{\oh },\,\tilde V^{\oh }$ and $\phi^{\oh },\,\tilde \phi^{\oh }$ 
belong to the $SU(2)$ and $SL(2)$ conformal field theories, respectively. 
Their weights are $\pm\oh $. Explicitly,
\begin{equation*}
  V^{\oh }(z)=\ep^{\frac{i}{\sqrt 2}\vaf (z)}
  \quad,\quad
  \tilde V^{\oh }(z)=\ep^{-\frac{i}{\sqrt 2}\vaf (z)} 
  \quad,\quad
  \phi^{\oh}(z)=\gamma\,\ep^{\frac{i}{\sqrt 2}\vas (z)}
  \quad,\quad
  \tilde\phi^{\oh}(z)=\beta\,\ep^{-\frac{i}{\sqrt 2}\vas (z)}
  \quad.
\end{equation*}

Before going on, it would be interesting to interpret the $N=4$
highest weight vectors \eqref{hwv-N4} as representations of the
underlying world-sheet $\so\times\su$ algebra. In thinking about this
question, a natural analogy arises with 2d gravity. This may be seen
explicitly as follows. In the $SU(2)$ conformal field theory (WZW
model at a fixed point) of level $1$, there are only two primary
fields which belong to the basic grid of the model, i.e. $j\leq\oh$.
Therefore, it is at first sight natural to expect that the $N=4$
primary fields are these $SU(2)$ fields dressed by proper $SL(2)$
fields. From our discussion of this issue, we have seen that there
are, however, extra states for $k>1$ appearing as dressed type states
outside the basic grid, i.e. $j>\oh\,$\footnote{Strictly speaking, the
  $\ep^{i\sqrt 2j\varphi}$ are not primary for the $SU(2)$ WZW model
  of level $1$.}. It is well-known that the same happened within 2d
gravity coupled to the minimal conformal matter.

We will here conclude this section with brief observations about
string theory on $\ads\times\text{S}^3$. It has been much studied
recently in both the NSR and GS formulations (see, e.g., [7-18, 30-32]
and refs. therein).  Here we will note interesting facts relevant to
this problem.

The starting point within the NSR approach to string theory on
$\ads\times\text{S}^3$ is first to consider the corresponding bosonic
theory.  This usually involves the $SL(2)\times SU(2)$ WZW model as a
keystone. To extend the analysis to superstrings, the Virasoro
(conformal) algebra on the world-sheet is replaced by the $N=1$
superconformal algebra. Technically, this is achieved by introducing
free fermions taking their values in the Lie algebra $sl(2)\times
su(2)$. The NSR framework makes it possible to discuss a lot but it is
not very convenient for discussing the space-time supersymmetry
because it requires working in all pictures simultaneously. In other
words, the algebra does not close on a finite number of Bose
sea-levels for the superconformal ghosts. What we have done so far is
just recall how to get the space-time supersymmetry in the NSR
description of $\ads\times\text{S}^3$.  Now we wish to show that there
exists a simpler formulation that results in the space-time
supersymmetry. In doing so, we must bear in mind that the bosonic
$SL(2)\times SU(2)$ WZW model is bosonized by the Wakimoto and vertex
operator constructions as given by \eqref{bN4} and \eqref{Wr}. Then
the $N=4$ superconformal generators are simply given by \eqref{qn4}!
Thus, we do not need the $N=1$ superconformal algebra anymore.
Moreover, we get a manifest supersymmetry which is due to the fermions
$\xi ,\txi $ hidden by such bosonization. Both of these features are
typical for the GS formulation.  Can we interpret this construction as
a hybrid of the NSR and GS formulations? We cannot rule out this
hypothesis because it is well-known that $D=2$ is a delicate case from
the conformal field theory point of view.  So, something special may
happen.


\section{Conclusions and Remarks}
\renewcommand{\theequation}{5.\arabic{equation}}
\setcounter{equation}{0}

First let us say a few words about the results.

In this work we have built the unitary representations of the affine
Lie algebras A, D, E type as well as the massless representations of
the $N=4$ superconformal algebra that is a natural continuation of
what we did in our previous work. It is found that the postulates
(i)-(iii), as these are formulated in the introduction, are not only
useful for string propagation on $\ads$, but are in fact crucial to
get new constructions. Indeed, using the Green's idea ``world-sheets
for world-sheets'' results in new free field representations of affine
Lie algebras that contain fewer free fields than the corresponding
Wakimoto representations have. The same is also true for the $N=4$
superconformal algebra. It opens new possibilities in studying
infinite dimensional Lie algebras. Moreover, we have presented the
hints in favor of the existence of a new construction (hybrid) for
string propagation on $\ads$. It is clear that further work is needed
to make this more rigorous.  So, let us conclude by mentioning a few
problems that are seemed the most important to us.

(i) Of course, the most important open problem is to understand how to
incorporate the massive representations of the $N=4$ superconformal
algebra into our construction. As we noted in subsection 4.1, these
representations exhibit a continuum in the spectrum of the conformal
dimensions. This is exactly the reason why our construction fails.
However, there exists a deeper problem. Up to now it is not known what
the OP algebra and correlation functions of the $N=4$ CFT are.

(ii) The new construction proposed in section 3 to get the free field
representations of affine Lie algebras of the A, D, E, type requires
much work in the future to reach such mathematical understanding as we
have for the vertex operator construction and the Wakimoto free field
representation. Our analysis of section 3 is essentially a physical
one.

(iii) Finally, the problem that obviously deserves more attention is
the hypothesis about a hybrid of the NSR and GS formulations for
string theory on $\ads$. Here we need a fuller understanding of the
problem on the level of a proper Polyakov path integral.

\vspace{.25cm} {\bf Acknowledgments}

\vspace{.25cm} It is a pleasure to thank B. Feigin and R. Metsaev for
many stimulating discussions, and A. Semikhatov for reading the manuscript 
and a very useful discussion.  This research was supported in part by
Russian Basic Research Foundation under grant 9901-01169.



\end{document}